\documentclass[11pt]{article} 
\usepackage{amsfonts,amscd,amsmath,graphicx}

\def\NP#1#2{Nucl.Phys. B#1 (#2)} 
\def\PL#1#2{Phys.Lett. B#1 (#2)}

\def\PR#1#2{Phys.Rev. D#1 (#2)} 
\def\IJMP#1#2{Int.J.Mod.Phys. A#1 (#2)}

\def\HP#1#2{JHEP #1 (#2)} 
\def\ap{ \alpha^{\prime}} 

\def\2p{2\pi\alpha^{\prime}}
\def\pd{\partial}
\def\cl{T_{\text{c}}}
\def\dil{\phi_{\text{c}}}
\def\Pc{\Phi_{\text{c}}}

\def\lae{\lambda_{e}}

\def\ar{a_{\R}}
\newcommand{\ep}{\text e}
\newcommand{\oh}{\frac{1}{2}}

\newcommand{\R}{\text{\tiny R}}
\title{ On One-Loop Approximation to Tachyon Potentials}
\author{Oleg Andreev\thanks{e-mail:  andreev@physik.hu-berlin.de}
\thanks{Permanent address: Landau Institute for Theoretical Physics, Moscow, Russia}
\hspace{.15cm} and 
Tassilo Ott\thanks{e-mail: ott@physik.hu-berlin.de}
\\ \\
Humboldt--Universit\"at zu Berlin, Institut f\"ur Physik\\
Invalidenstra\ss e 110, D-10115 Berlin, Germany}

\date{}
\begin{document} 
 \maketitle 
\begin{abstract} 
We compute the one-loop corrections to the tachyon potentials in both, bosonic and supersymmetric, cases 
by worldsheet methods. In the process, we also get some 
insight on tachyon condensation from the viewpoint of closed string modes and show that there is an instability 
due to the closed string tachyon in the bosonic case. 
\end{abstract}

\vspace{-10.5cm}
\begin{flushright}
hep-th/0109187      \\
HU Berlin-EP-01/32
\end{flushright}
\vspace{9.5cm}

%\vspace{-5cm}
%_______________________     I N T R O D U C T I O N_________________
\section{Introduction}
\renewcommand{\theequation}{1.\arabic{equation}}
\setcounter{equation}{0}
It is a big problem to better understand the vacuum structure of string/M theory. Even in the case of bosonic 
string theory which contains the tachyon near its perturbative vacuum, it turned out not to be a simple deal to find 
a stable 
vacuum. Note that hopes for the existence of such a vacuum have their origins already in the seventies \cite{bar}. 
However, for open strings a real breakthrough has been achieved in recent years  when the open string tachyons 
were related with annihilation or decay of D-branes via the process of tachyon 
condensation \cite{as} (see \cite{rev} for a review and a list of references). 

In studying the phenomenon of tachyon condensation, string field theory methods are the most appropriate ones 
(see \cite{w-z} for the latest developments). However, in practice it turns out to be very hard to deal with them. 
This was a motivation for finding simpler methods, for example toy models \cite{z}, that can allow one to get some 
intuition on the physics  of the phenomenon. Recently, it was realized that the background independent open 
string field theory (BIOSFT) of Witten \cite{W} is a powerful tool to attack the problem. In particular, it 
allows one to compute the tree level effective tachyon potentials \cite{gera,kud,kud1} which amusingly 
coincide with the potentials found from the toy models based on the exactly solvable Schr\" odinger problem \cite{z}. 

The background independent open string field theory is based on the Batalin-Vilkovisky formalism \cite{bv} whose 
master equation provides the effective action of the theory. In the bosonic case, the classical master equation 
gives a solution \cite{W}
\begin{equation}\label{wd}
S_0=\left(1+\beta^i\frac{\pd}{\pd g^i}\right)Z_0
\quad,
\end{equation}
where $Z_0$ is the renormalized partition function of the underlying two-dimensional theory defined on 
the unit disk in the 
complex plane in such a way that it is a free theory in the bulk but an interacting theory on the boundary. 
So, the $g^i$ stand for coupling constants of boundary interactions and $\beta_i$ for their RG $\beta$-functions. 

In the supersymmetric case \footnote{We mean worldsheet SUSY, here and below.}, the solution of the classical 
master equation is simply \cite{super}
\begin{equation}\label{pkud}
S_0=Z_0
\quad.
\end{equation}

As we have already mentioned, it is very simple to compute the tree level effective tachyon potentials from 
formulae \eqref{wd} and \eqref{pkud}. However, to investigate the vacuum structure of the theory, it 
is necessary to have information about quantum corrections to the potentials. Unfortunately, the quantum master 
equation which is a proper machinery is still missing for the BIOSFT. The main problem here is a poor 
understanding of the space of boundary interactions of the underlying two-dimensional theory. 

However, we think that there are enough tools at our disposal to shed some light onto the problem of interest. Indeed, 
having the tree level effective actions, we can evaluate the loop corrections to the tachyon potentials by 
using ordinary field theory methods. On the other hand, following the BIOSFT, we can also compute partition 
functions of open strings in the presence of a non-zero tachyon background on Riemann surfaces which 
correspond to string loops. In general, the appearance of these surfaces is not surprising as the effective action 
has to reduce to the partition function on-shell. In doing so, the comparison with the field theory results is crucial. 
Then we need to suggest how the loop effective actions are expressed in terms of the partition functions.  At the present 
time is not known whether the situation may be taken 
under control. So we are bound to learn something if we succeed. In what follows, we will restrict 
ourselves to the one-loop approximation that 
significantly simplifies the problem. Thus we will consider the partition functions on Riemann surfaces of zero genus 
in the presence of  a constant tachyon background.

We should also mention that there have been attempts to generalize the tree level machinery of the BIOSFT to 
next order \cite{1b,1s} \footnote{For a recent work on quantum corrections 
within $p$-adic string theory see, \cite{min}.}. However, they did not discuss the consistency with 
the field theory results. As a consequence, it has not become completely clear whether tachyon condensation 
described by the  BIOSFT leads to the strong coupling regime as the field theory results assume or 
not \footnote{For a discussion of this issue with further references see, e.g., \cite{s}.}.

The outline of the paper is as follows. We start in section 2 by directly evaluating the one-loop corrections to the 
tachyon potentials based on the toy field theory models as well as the tree level actions provided by the BIOSFT. 
We do so for both cases, the bosonic and supersymmetric ones. We continue to pursue our program in section 3 by 
evaluating the partition functions of open strings with a constant tachyon background on Riemann surfaces of 
zero genus. In doing so, the comparison with the field theory results is crucial. In section 4, we define the effective 
action via the partition function and find the one-loop corrections to the tree level tachyon potentials. We will also 
discuss the phenomenon of tachyon condensation as it is seen by closed strings. The picture that emerges seems 
quite curious as in terms of closed strings, tachyon condensation looks like 
the splitting of an original D-brane into two copies which are running apart while the tachyon is rolling down 
to the local minimum. Section 5 will present our conclusions and directions for future work. In the appendix we 
give some technical details which are relevant for our discussion in sections 3 and 4.

%____________________________________________________________________________
\section{Effective Field Theory Analysis}
\renewcommand{\theequation}{2.\arabic{equation}}
\setcounter{equation}{0}
The goal of the present section is to evaluate the one-loop corrections
to the tree level potentials based on some field theory models of unstable branes. Our discussion in this section is 
elementary. We begin with the bosonic string.
\subsection{Toy Models for D-branes of Bosonic String Theory}
The field theory model of \cite{z} as well as the $\text{p}\rightarrow 1$ limit of the tree level effective action of 
the tachyon field in p-adic string theory \cite{gera} lead to the following action on the unstable bosonic brane 
\begin{equation}\label{p}
\text{S}_0=\int\,d^dx\,\,\Bigl[
\oh\ap\,\pd_i\Phi\pd_i\Phi+\frac{1}{4}\Phi^2\left(1-2\ln\Phi\right)\Bigr]
\quad.
\end{equation}
From now, we use the Minkowski metric on the D-brane. Thus, the metric for the transverse directions 
is always euclidean.

A formal evaluation of the one-loop correction to the potential results in \footnote{See, e.g., \cite{jean}.}
\begin{equation}\label{p1}
V_{1-loop}=-\oh\int_0^{\infty}\frac{dl}{l}\bigl(2\pi l\bigr)^{-\frac{d}{2}}\,
\ep^{-\oh \text{m}_{\text{e}}^2l}
+
\oh\int_0^{\infty}\frac{dl}{l}\bigl(2\pi l\bigr)^{-\frac{d}{2}}\,
\ep^{-\oh \text{m}_0^2l}
+P(\Pc)
\end{equation}
with 
\begin{equation}\label{m1}
\text{m}_{\text{e}}^2= -\frac{1}{\ap}\bigl(1+\ln\Pc\bigr)
\quad.
\end{equation}
The second term is a normalization, so $\text{m}_0^2$ equals $-1/\ap$ or $\infty$. $P(\Pc)$ stands for 
possible counterterms which are fixed by the renormalization conditions. 

So far we have just given a model which could be used in principle as a toy model for tachyon condensation. However, it 
suffers from some drawbacks: for example, it does not reproduces the descent relations for brane tensions. Because of 
this, let us consider the effective action that comes out from the background independent open string field 
theory \cite{W}. Restricting to $\pd^2$ order, the action of the tachyon field on the bosonic Dp-brane 
takes the form \cite{gera,kud}
\begin{equation}\label{sm}
S_0={\text T}_p\int\,d^dx\,\,\ep^{-\bigl(T+\dil\bigr)}
\Bigl[1+T+\ap\,\pd_iT\pd_iT+O\left(\pd^4\right)\Bigr]
\quad,
\end{equation}
where ${\text T}_p$ is the tension of the $\text{D}_p$-brane and $p=d-1$. We also 
include a constant factor of the dilaton. So, $\dil$ means its vacuum expectation value. Note also that modulo 
a numerical coefficient in front of the kinetic term, one can get one model from the other by a relation 
$\Phi=\ep^{-\oh T}$. 

Like in the previous case, the one-loop correction to the potential can be written  in the form \eqref{p1} 
but now with 
\begin{equation}\label{m2}
\text{m}^2_{\text{e}}\rightarrow m_e^2= \frac{1}{2\ap}\bigl(\cl -1\bigr)
\quad.
\end{equation}
As we see, the tachyon mass near the open string perturbative vacuum ($T=0$) differs from its correct value. 
This may sound like a problem, but in fact it is not, because the expansion in $\pd$ as it is used in Eq.\eqref{sm} fails 
near this vacuum. So one has to use the expansion in $T$  or, alternatively, to take into account higher derivative 
terms \cite{kud}. It is clear that the one-loop correction to the potential receives contributions from such terms, so our 
result \eqref{m2} will receive corrections too. According to the results of \cite{kud}, the tachyon mass near the open 
string perturbative vacuum is finally getting corrected in a proper way. Since in both cases we have the same dependence 
on the background field (modulo a field redefinition), it is natural to assume that the right hand side of Eq.\eqref{m2} 
is only corrected by $-1/2\ap$.

Now we will make some remarks about these models:
\newline (i) In the problem of interest the perturbation theory coupling constant is given by 
\begin{equation}\label{cc}
\lae\sim\ep^{\oh\bigl(\cl +\dil\bigr)}
\quad,
\end{equation}
that allows one to say that the theory is getting strongly coupled near the vacuum $T=\infty$ (see, e.g., \cite{s} and 
refs. therein). By mapping 
the field $T$ into $\Phi$ as $\Phi\sim\ep^{-\oh T}$, there seems to be no way to ``escape'' the strong coupling 
regime in the vicinity of this vacuum without coming back to p-adic string theory with a general parameter p.
\newline(ii) It immediately follows from \eqref{cc} that we can trust our one-loop estimation at least for a 
range \footnote{We assume that in the open string perturbative vacuum the theory is 
weakly coupled, i.e. $\dil\ll 0$.}
\begin{equation}\label{range}
\cl <\vert\dil\vert
\quad.
\end{equation}
\newline(iii) Since the field theory analysis makes no sense for $d=0$, we do not consider D-instantons but only 
comment on them in section 5.  
 
%____________________________________________________________________________
\subsection{Toy Models for Unstable D-branes of  Superstring Theory}
Now we want to briefly discuss what the field theory actions of unstable D-brane in superstring theory 
look like. We consider a generic case, with the worldvolume of a brane parameterized by $x^1,\dots,x^d$. 
Thus, we assume that the even and odd $d$'s are referred to type IIA and IIB, respectively.

The field theory model of \cite{z} is given by 
\begin{equation}\label{zs}
\text{S}_0=\int\,d^dx\,\,\Bigl[
\oh\ap\,\pd_i\Phi\pd_i\Phi+
\frac{1}{2\pi}\ep^{-2\bigl(\text{erf}^{\,-1}(\Phi)\bigr)^2}
\Bigr]
\quad.
\end{equation}
where 'erf' stands for the error function: $\text{erf}(x)=\frac{2}{\sqrt{\pi}}\int_0^x dz\,\ep^{-z^2}$. 

The one-loop correction to the potential is given by \eqref{p1} with the effective mass 
\begin{equation}\label{ss1}
\text{m}^2_{\text{e}}\rightarrow 
\text{M}_{\text{e}}^2= \frac{1}{2\ap}
\Bigl(2\Bigl[\text{erf}^{\,-1}\bigl(\Pc\bigr)\Bigr]^2-1\Bigr)
\quad.
\end{equation}

On the other hand, the background independent open string field theory provides the following action 
of the tachyon field on the unstable Dp-brane \cite{kud1}
\begin{equation}\label{ss}
S_0={\text T}_p\int\,d^dx\,\,\ep^{-\bigl(T^2+\dil\bigr)}
\Bigl[1+4\ln2\,\ap\,\pd_iT\pd_iT+O\left(\pd^4\right)\Bigr]
\quad.
\end{equation}
 The correction to the potential again is given by the general form \eqref{p1} but with the effective mass 
\begin{equation}\label{m3}
\text{m}^2_{\text{e}}\rightarrow 
M_e^2= \frac{1}{2\ln2\ap}\Bigl(\cl^2-\oh\Bigr)
\quad.
\end{equation}
Just as in the bosonic case, this BIOSFT model does not give the correct mass for the tachyon field 
near the perturbative vacuum. The reason are again higher derivative corrections in \eqref{ss}. Note that modulo 
a numerical coefficient in front of the kinetic term, one can get the first model by  a relation 
$\Phi=\text{erf}\bigl(T/\sqrt{2}\bigr)$. 

Finally, let us note that the effective coupling constant of perturbation theory is 
\begin{equation}\label{ccs}
\lae\sim\ep^{\oh\bigl(\cl ^2+\dil\bigr)}
\quad
\end{equation}
that allows us to stay at the weak coupling regime as far as 
\begin{equation}\label{ranges}
\cl^2 <\vert\dil\vert
\quad.
\end{equation}

%____________________________________________________________________________
\section{World-Sheet Analysis}
\renewcommand{\theequation}{3.\arabic{equation}}
\setcounter{equation}{0}
In this section we consider open strings in the presence of a constant tachyon background. Our primary interest 
is to compute the partition functions on a world-sheet with Euler number zero. Since we are interested in the 
oriented theories, we restrict ourselves to the annulus and cylinder.
%_________________________________________________________________________________
\subsection{Some Computations for Bosonic String}
Before going to explicit computations, let us stress a couple of important points:
\newline (i) We will consider boundary interactions due to non-primary conformal fields, so the use of 
conformal maps to transform one worldsheet into another one is pointless because the transformation laws of the 
fields are unknown. Note that constant background fields are defined via the Taylor expansion 
in $X$ of $X$-dependent backgrounds, as for example in the BIOSFT, so they can also transform in a non-trivial 
way under the maps.
\newline(ii) It is clear that the physics in spacetime must be universal in a sense that it does not depend on whether 
one uses the cylinder or annulus. As we have noted in the problem of interest the use of conformal 
machinery is problematic, so we have to involve other principles. A related problem is that there are infinitely 
many non-equivalent backgrounds that makes the computation of the partition function ambiguous. To fix these 
problems, we propose to use a comparison with field theory results as basic principle. Indeed,  given the partition 
function, one can in principle extract the contribution due to the tachyon from it and fix the background by comparison 
with the corresponding field theory result computed from the tree level effective action. As we will see shortly, this also leads 
to equivalent expressions for the partition functions computed for the annulus and cylinder.

\subsubsection{Computations Using Annulus}

First we would like to see what happens on a world-sheet whose geometry is an annulus. A motivation for doing 
so comes from the fact that the BIOSFT is defined on the disk, so it seems natural to expect that one-loop corrections 
come from the theory on the annulus. Moreover, there is another interesting problem which remains to be fixed: the 
field theory analysis results in the effective coupling constant \eqref{cc} which assumes that the theory becomes 
strongly coupled due to tachyon condensation while the world-sheet analysis of \cite{har,kud} leads to an opposite 
conclusion. 

To proceed, we briefly recall the main points of the theory on a disk \cite{W}. We add the boundary interaction
to the standard action. 
\begin{equation}\label{bon1}
S_b^{disk}=\frac{1}{2\pi}\int rd\theta \,T(X)
\quad,
\end{equation}
where $r$ is the radius. In what follows, we restrict ourselves to a constant value for the tachyon 
background, i.e. $T(X)=\mathbf a$.  

The computation of the tree level effective action along the lines of the BIOSFT is elementary. First, we define a 
renormalized coupling $\mathbf{a}_{\text{\tiny R}}=r\mathbf{a}$ that results in the following $\beta$-function 
$\beta_{\mathbf{a}_{\text{\tiny R}}}=-\mathbf{a}_{\text{\tiny R}}$. Next, using the fact that the renormalized 
partition function on the disk is $\ep^{-\mathbf{a}_{\text{\tiny R}}}$, we get from the formula \eqref{wd} the same 
potential as in \eqref{sm} after the identification $\cl=\mathbf{a}_{\text{\tiny R}}$. 

It is also straightforward to compute the partition function on the annulus whose inner and outer radii are $qr$ and 
$r$, respectively. In this case we add two tachyon backgrounds 
\begin{equation}\label{bon2}
S_b^{ann}=\frac{1}{2\pi}\int rd\theta \,\Bigl[q\,T^{in}(X)+T^{out}(X)\Bigl]
\quad
\end{equation}
and restrict ourselves to constant backgrounds: $T^{in}=\mathbf{a}^{in}_{\text{\tiny R}}$ and 
$T^{out}=\mathbf{a}^{out}_{\text{\tiny R}}$. We define renormalized couplings: 
$\mathbf{a}^{in}_{\text{\tiny R}}=r\mathbf{a}^{in}$ and $\mathbf{a}^{out}_{\text{\tiny R}}=r\mathbf{a}^{out}$.

Since the boundary interaction does not depend on $X$, the computation of the corresponding path integral 
is elementary. We get
\begin{equation}\label{pf-ann}
Z_1^{ann}(\mathbf{a}^{out}_{\text{\tiny R}},\mathbf{a}^{in}_{\text{\tiny R}})
=i\oh V_d \bigl(8\pi^3\ap)^{-\frac{d}{2}}\pi^{12}
\int_0^1\frac{dq}{q^3}\,\bigl(-\ln q\bigr)^{\frac{d}{2}-13}
\ep^{-\mathbf{a}^{out}_{\text{\tiny R}}-q\,\mathbf{a}^{in}_{\text{\tiny R}}}\,
\prod_{n=1}^{\infty}\bigl(1-q^{2n}\bigr)^{-24}
\quad,
\end{equation}
where $V_d$ is the D-brane volume.

It is natural to assume that $\mathbf{a}^{out}_{\text{\tiny R}}$ coincides with the corresponding background on 
the disk, i.e. $\mathbf{a}^{out}_{\text{\tiny R}}=\mathbf{a}_{\text{\tiny R}}=\cl$. In other words, we consider 
the annulus as being produced from the disk by cutting a hole. Although this procedure has no influence on the 
background at the outer boundary, it does not say how to define the background at the new inner boundary. The later 
makes the partition function ambiguous.  As we have already mentioned, a conformal map which maps one boundary 
into one another does not help as the boundary interactions are not due to the primary conformal operators and their 
transformation laws are unknown. To fix the background, we use the comparison with the field theory result of 
section 2. Up to an overall factor, the contribution due to the open string tachyon is given by 
\begin{equation}\label{pf-ann1}
\int_0^1\frac{dq}{q}\,\bigl(-\ln q\bigr)^{\frac{d}{2}-1}
\ep^{-\mathbf{a}_{\text{\tiny R}}-q\,\mathbf{a}^{in}_{\text{\tiny R}}}\,
\ep^{-\frac{2\pi^2}{\ln q}}
\quad.
\end{equation}
Using the fact that $\mathbf{a}_{\text{\tiny R}}=\cl$, we find $\mathbf{a}^{in}_{\text{\tiny R}}$ by comparing 
the exponents in \eqref{pf-ann1} and \eqref{p1}
\begin{equation}\label{in1}
\mathbf{a}^{in}_{\text{\tiny R}}=-\frac{1}{q}\mathbf{a}_{\text{\tiny R}}
-\frac{\pi^2}{q\ln q}\,\mathbf{a}_{\text{\tiny R}}
\quad.
\end{equation}
Finally, we get 
\begin{equation}\label{pf-ann2}
Z_1^{ann}(\mathbf{a}_{\text{\tiny R}})
=i\oh V_d \bigl(8\pi^3\ap)^{-\frac{d}{2}}\pi^{12}
\int_0^1\frac{dq}{q^3}\,\bigl(-\ln q\bigr)^{\frac{d}{2}-13}
\ep^{\frac{\pi^2}{\ln q}\mathbf{a}_{\text{\tiny R}}}\,
\prod_{n=1}^{\infty}\bigl(1-q^{2n}\bigr)^{-24}
\quad.
\end{equation}

\subsubsection{Computations Using Cylinder}

At first glance, the relation between the backgrounds on the different boundaries might seem strange. However, 
it is the price to pay for the unknown transformation law of the background fields under the conformal map 
of one boundary into another. From this point of view it seems reasonable to consider the cylindric geometry when 
both boundaries are apparently treated on an equal footing, so the backgrounds on them are the same. Moreover, 
the cylinder is a finite strip obtained by cutting from an infinite one. In doing so, one does not change the background 
field living on the boundaries. Thus the background is fixed at the tree level by a comparison with field theory. 
So, the computation of the partition function may be considered as a consistency check. 

To proceed, we consider the open bosonic string on the  worldsheet parameterized by 
coordinates $(\tau,\sigma)$ with a range for the spacial coordinate $\sigma$ to be $0\leq\sigma\leq\pi$. 
We add the boundary interaction 
\begin{equation}\label{bon}
S_b=\frac{1}{4}\int d\tau d\sigma \Bigl[T(X)\delta(\sigma)+T(X)\delta(\pi-\sigma)\Bigr]
\quad
\end{equation}
to the standard action and restrict ourselves to a constant tachyon background (coupling constant) $T=a$ in 
what follows.

The tree level effective action is easily computed via \eqref{wd} treated now as an ansatz. First, we 
introduce a scale in the problem at hand that restricts $\tau$ to $\vert\tau\vert\leq r$ and simply rescales 
$\sigma$. Next, we define a renormalized coupling as $\ar=ra$. This implies that the renormalized partition function 
on the strip is simply $Z_0=\ep^{-\ar}$ and the RG beta function is $\beta_{\ar}=-r\frac{\pd}{\pd r}\ar=-\ar$. 
Finally, a simple algebra results in the same potential as in \eqref{sm} after the identification $\ar=\cl$.

Having fixed the background, let us now compute the partition function on the cylinder. Again, since the boundary 
interaction does not depend on the $X$'s, the path integral is easily worked out. The result is given by
\begin{equation}\label{pfc1}
Z_1(\ar)=i\oh V_d\int_0^\infty\frac{dt}{t}\,\bigl(8\pi^2\ap t\bigr)^{-\frac{d}{2}}\ep^{-\pi t\ar}\,
\eta(it)^{-24}
\quad,
\end{equation}
where $\eta$ is the Dedekind eta function.

As a consistency check, we extract from $Z_1(\ar)$ the contribution due to the open string tachyon 
\begin{equation}\label{check}
\int_0^\infty dt\,t^{-1-\frac{d}{2}}\,
\ep^{(2-\ar)\pi t}
\quad.
\end{equation}
A comparison with the field theory result of section 2 gives $\ar=\cl$. Here we have found the same relation between 
the renormalized value of the tachyon background and the classical field as at the tree level that is of course 
a desired result. Moreover, a simple algebra shows that the expression for the partition function we have found on the 
cylinder coincides with the one on the annulus. This is again a desired result. Note that we have not used the 
conformal map for mapping the annulus into the cylinder from the beginning. The reason for doing so is that 
we do not know the transformation law for the background fields.

At this point a comment is in order. We should note that in the case of a constant tachyon background the 
representation of the partition function as a loop of open string states is much like that without any background. 
Indeed, one can rewrite it 
as \footnote{For $\ar=0$, see \cite{dj}.} 
\begin{equation}\label{particle}
Z_1(\ar)=i\oh V_d\sum_{n=0}^\infty d_n\int_0^\infty\frac{dl}{l}\,\bigl(2\pi l\bigr)^{-\frac{d}{2}}
\ep^{-\oh l\text{m}^2_{\text{e}}(n)}\,
\quad,
\end{equation}
where 
\begin{equation}\label{mass}
\text{m}^2_{\text{e}}(n)=\frac{1}{\ap}\Bigl(n-1+\frac{\ar}{2}\Bigr)
\end{equation}
and $d_n$ is defined as the coefficient 
of $q^n$ in $\prod_{n=1}^\infty\bigl(1-q^n\bigr)^{-24}$. This representation looks like a sum of 
the point-particle free energies (in the Coleman-Weinberg representation) over the open string spectrum but 
with the masses of the particles shifted according to Eq.\eqref{mass}. It is natural to treat the particle with 
$n=0$ as the tachyon and compare its contribution with the field theory result. In fact, 
the last assumes that the partition function coincides with the one-loop correction to the effective action. We will 
return to this point in section 4. 

\subsubsection{Partition Function Analysis}

Now, let us analyze the expression for the partition function. Be reminded what happens for $\ar=0$:
In this case  the integrand has divergencies as $t\rightarrow\infty$ or $t\rightarrow 0$ which come from 
the lightest string modes. In order to see what is going on in the case of a nonzero tachyon background, we 
begin with $t\rightarrow\infty$ . In studying this limit, we really need the representation of 
the partition function in terms of a loop of open string modes, namely Eq.\eqref{pfc1} (or equivalently \eqref{particle}). 
Expanding the products in $e^{-2\pi t}$, we get
\begin{equation}\label{inf}
Z_1(\ar)=i\oh V_d\int_0^\infty\frac{dt}{t}\,\bigl(8\pi^2\ap t\bigr)^{-\frac{d}{2}}
\ep^{(2-\ar)\pi t}\,
\Bigl[1+24\,\ep^{-2\pi t}+\dots\Bigr]
\quad.
\end{equation}
In the range \eqref{range} ($\ar=\cl$), a possible divergence can only come from the leading term if  
the value of the background is less than its critical value $a_0=2$. It can be fixed by using the analytic 
continuation in $\ar$. Indeed, evaluating the integral for $\ar>2$ as
\begin{equation}\label{int}
\int_0^\infty\,dt\,t^{-1-\frac{d}{2}}\ep^{-(\ar-2)\pi t}=\bigl[\pi(\ar-2)\bigr]^{\frac{d}{2}}\,
\Gamma(-\frac{d}{2})
\end{equation}
and  then defining it for $\ar<2$ via the analytic continuation 
$(\ar-2)^{\frac{d}{2}}=\ep^{\pm i\pi \frac{d}{2}}\vert\ar-2\vert^{\frac{d}{2}}$, we get the desired result. At 
this time we do not care about the last factor and the sign in $\ep^{\pm i\pi \frac{d}{2}}$.

To complete the picture, let us also evaluate the contribution of massive open string modes in Eq.\eqref{inf}. 
Integrating over $t$, we end up with the following sum
\begin{equation}\label{om}
\sum_{n=1}^{\infty} d_{n+1}\,\bigl(\ar+2n\bigr)^{\frac{d}{2}}
\quad,
\end{equation}
It is well-known that $d_n$ rapidly grows with $n$ that  leads to a divergency. To be more precise, 
for $n\rightarrow\infty$, $d_n$ is asymptotically \footnote{See, e.g., \cite{gsw}.}
\begin{equation}\label{d}
d_n\sim n^{-27/4}\,\ep^{4\pi\sqrt{n}}
\quad.
\end{equation}
Thus, the contribution of massive open string modes is divergent and this is the case for any value of the 
background field. We will return to this point shortly after working out the other limit.

We can readily rewrite the product in \eqref{pfc1} in terms of $\ep^{-\frac{2\pi}{t}}$. This yields 
another representation of the partition function which is useful in studying the limit $t\rightarrow 0$. 
We then get the following expansion of the partition function 
\begin{equation}\label{inf2}
Z_1(\ar)=i\oh V_d\int_0^\infty\frac{dt}{t^{-11}}\,\bigl(8\pi^2\ap t\bigr)^{-\frac{d}{2}}
\ep^{(\frac{2}{t}-\ar t)\pi}\,
\Bigl[1+24\,\ep^{-\frac{2\pi}{ t}}+\dots\Bigr]
\quad.
\end{equation}

Before continuing our discussion, we will make a detour and recall some basic results on the 
Feynman propagator of a scalar particle of mass $m$ in $d$ dimensions that is given by the familiar formula 
\begin{equation}\label{fpr}
G_d\bigl(x,x';m^2\bigr)=\frac{1}{(2\pi)^d}\int d^dp\,\frac{\ep^{ip(x-x')}}{p^2+m^2}
\quad
\end{equation}
and can also be rewritten as 
\begin{equation}\label{fpr1}
G_d\bigl(R;m^2\bigr)=\frac{1}{(4\pi)^{\frac{d}{2}}}\int_0^{\infty}dl\,
l^{-\frac{d}{2}}
\,\ep^{-m^2l-\frac{1}{4}\frac{R^2}{l}}
\quad,
\end{equation}
with $R=\sqrt{(x-x')^2}$. 

Given the integral representation for the propagator \eqref{fpr1}, we can readily rewrite the 
expansion \eqref{inf2} as 
\begin{equation}\label{inf3}
Z_1(\ar)=iV_d\,(4\pi^2\ap)^{12-d}\,\frac{\pi}{2^{11}}
\Bigl[G_{26-d}\bigl(R;-4/\ap\bigr)+24\,G_{26-d}\bigl(R;0\bigr)+\dots\Bigr]
\quad,
\end{equation}
where $R=\sqrt{2\pi^2\ap\ar}$. At this point a couple of remarks are in order.
\newline (i) Such a representation makes sense only in the case of $\ar\geq 0$. So, we will 
always assume this if not stated otherwise.
\newline (ii) In contrast with the representation in terms of open string modes (see \eqref{particle}), the 
representation \eqref{inf3} includes closed string modes whose masses are not shifted. 

As in the case of the representation via open string modes, a natural question to pose is 
whether the sum over all massive closed string modes is convergent or not. To proceed, we will use the 
familiar asymptotics for the propagator of a massive scalar particle 
\begin{equation}\label{fpr3}
G_d\bigl(R;m^2\bigr)=\frac{1}{2m}\Bigl(\frac{2\pi R}{m}\Bigr)^{\frac{1-d}{2}}\ep^{-mR}
\Bigl[1+O\Bigl(\frac{1}{mR}\Bigr)\Bigr]
\quad,\quad
mR\rightarrow\infty
\end{equation}
together with the asymptotics for $d_n$ that allows us to evaluate the contribution of the massive 
closed string modes in Eq.\eqref{inf3}
\begin{equation}\label{inf4}
\sum_{n=1}^\infty d_{n+1}\,G_{26-d}\bigl(R;4n/\ap \bigr)
\sim
\sum_{n=1}^\infty n^{-1-d/4}\,
\ep^{4\pi\sqrt{n}\bigl(1-\sqrt{\ar/2}\bigr)}
\quad.
\end{equation}
It is obvious that the sum is convergent if $\ar>2$, i.e. the background field must be larger than its critical value. 
Thus, the critical value $a_0$ now means convergence of not only a single contribution to the partition function 
by the open string tachyon but all massive closed string states \footnote{This reminds the situation where 
the partition function is ill-defined beyond a maximum temperature due to a high degeneracy of levels.}.

It is instructive to see if Eq.\eqref{inf4} provides the same scaling behavior near the critical point as 
Eq.\eqref{int}. Using the fact that in the vicinity of the critical value 
$1-\sqrt{\frac{\ar}{2}}=-\frac{1}{4}\bigl(\ar-2\bigr)+O\Bigl(\bigl(\ar-2\bigr)^2\Bigr)$, a 
simple estimation results in the desired scaling
\begin{equation}\label{inf5}
\sum_{n=1}^{\infty} n^{-1-d/4}\,\ep^{-\pi\bigl(\ar-2\bigr)\sqrt{n}}
\sim \bigl(\ar-2\bigr)^{\frac{d}{2}}
\quad.
\end{equation}
Having related the divergence due to the massive closed string modes to the open string tachyon, 
it seems natural to relate the divergence due to the closed string tachyon to the massive open string 
modes. As we have seen, the divergence due to the massive open string modes does not crucially depend on 
the value of the background field. That is also clear in terms of the closed string tachyon where it is due to 
the poles of the propagator.

We will now use similar methods to estimate the asymptotic behavior $Z_1$ for 
large $\ar$ \footnote{So, $\ar\gg a_0$.}. In this limit, the propagator of a massive 
mode is given by \eqref{fpr3} which allows us to estimate the contribution of all massive modes except the tachyon. 
It is given by the lightest massive mode and therefore exhibits an exponential falling. On the other hand, 
the contribution of massless modes which is due to the propagator 
\begin{equation}\label{fpr4}
G_d\bigl(R;0\bigr)=\frac{1}{4}\Bigl(\frac{1}{\pi}\Bigr)^{\frac{d}{2}}
\Gamma(\frac{d}{2}-1)\,R^{2-d}
\quad
\end{equation}
shows a power like behavior \footnote{Note that the two cases $d=0$ and $d=2$ deserve a special consideration. We 
will discuss this issue in section 4.}. This means that in this limit we can 
omit all massive modes which is the expected result. However, there is also the contribution of the closed 
string tachyon which makes the estimation more involved. In this case the propagator becomes 
divergent as it is clear from Eq.\eqref{fpr} or Eq.\eqref{fpr1}. So, one must somehow treat this divergency but then 
the final result is ambiguous, i.e., it depends on the way of treatment. We refer to the appendix for more 
details. Due to this, finding the asymptotics of the partition function for large $\ar$ is problematic. Moreover, 
it also becomes difficult to extract the real part of $Z_1$. Indeed, there are two contributions to $\Re Z_1$:
The first one comes from the open string tachyon or, equivalently, from massive closed string modes if $\ar <a_0$. 
It can be readily found from Eq.\eqref{int}. Explicitly, 
\begin{equation}\label{rez1}
\pm\frac{\pi}{2}V_d\,(8\pi\ap)^{-d/2}
\frac{\bigl(2-\ar\bigr)^{\frac{d}{2}}}{\Gamma(1+d/2)}H(2-\ar)
\quad,
\end{equation}
where $H(x)$ is the Heaviside step function. The second comes from massive open string modes, or 
equivalently, from the closed string tachyon propagator. So, from \eqref{inf3} we have
\begin{equation}\label{rez2}
-V_d(4\pi^2\ap)^{12-d}\,\frac{\pi}{2^{11}}\,\Im\,G_{26-d}\bigl(\sqrt{2\pi^2\ap\ar};-4/\ap\bigr)
\quad.
\end{equation}
We will return to the ambiguities by a discussion of physics in the next section.

%______________________________________________________________________________
\subsection{Some Computations for NSR String}
We will now use similar methods to compute the partition functions for the open NSR string in the presence of a 
constant tachyon background. Since the generalization of the results of section 3.1 is rather straightforward, we omit 
some technical details.

\subsubsection{Computations Using Annulus}

Let us first recall the computation within the open NSR string on the disk. So, we add the boundary interaction 
\begin{equation}\label{bons1}
S_b^{disk}=\frac{1}{2\pi}\int rd\theta \,T^2(X)
\quad
\end{equation}
to the standard worldsheet action. As in the bosonic case, we consider a constant background field $T=\mathbf{a}$. 

In computing the partition function on the disk, the only novelty compared to the bosonic computation is that 
the renormalized coupling is now defined as $\mathbf{a}_{\text{\tiny R}} =\sqrt{r}\mathbf{a}$. This leads to a 
partition function $Z_0^{disk}=\ep^{-\mathbf{a}_{\text{\tiny R}}^2}$ that together with the definition 
\eqref{pkud} provides the same potential as in \eqref{ss} after the identification $\mathbf{a}_{\text{\tiny R}}=\cl$. 
Note that the R sector gives no net contribution to $Z_0^{disk}$. 

Having computed the partition function on the disk, we now want to consider the partition function on the annulus. We 
proceed in the same way as in the bosonic case namely, we introduce two backgrounds
\begin{equation}\label{bons2}
S_b^{ann}=\frac{1}{2\pi}\int rd\theta \,\Bigl[q\,\bigl(T^{in}(X)\bigr)^2+\bigl(T^{out}(X)\bigr)^2\Bigl]
\quad
\end{equation}
and restrict ourselves to constant backgrounds: $T^{in}=\mathbf{a}^{in}_{\text{\tiny R}}$ and 
$T^{out}=\mathbf{a}^{out}_{\text{\tiny R}}$. We define renormalized couplings: 
$\mathbf{a}^{in}_{\text{\tiny R}}=\sqrt{r}\mathbf{a}^{in}$ and $\mathbf{a}^{out}_{\text{\tiny R}}=
\sqrt{r}\mathbf{a}^{out}$. 

In the case of constant backgrounds the computation of the partition function is a simple task. We find
\begin{equation}\label{pf-anns}
\begin{split}
Z_1^{ann}(\mathbf{a}^{out}_{\text{\tiny R}},\mathbf{a}^{in}_{\text{\tiny R}})
=i\oh V_d \bigl(8\pi^3\ap)^{-\frac{d}{2}}\pi^4
\int_0^1
&\frac{dq}{q}\,\bigl(-\ln q\bigr)^{\frac{d}{2}-5}
\ep^{-(\mathbf{a}^{out}_{\text{\tiny R}})^2-q(\mathbf{a}^{in}_{\text{\tiny R}})^2}\,
\prod_{n=1}^{\infty}\bigl(1-q^{2n}\bigr)^{-8}
\\
&\times\frac{1}{q}
\Bigl[\prod_{n=1}^{\infty}\bigl(1+q^{2n-1}\bigr)^8-
\prod_{n=1}^{\infty}\bigl(1-q^{2n-1}\bigr)^8\Bigr]
\quad.
\end{split}
\end{equation}
Note that in terms of the closed string modes, the partition function is expressed only via the NS-NS states. This sounds 
good because we are interested in unstable branes. Moreover, it also assumes that there is no closed string tachyon 
because of the corresponding cancellation between the NS and R sectors. 

As in the bosonic case, it is natural to assume that $\mathbf{a}^{out}_{\text{\tiny R}}$ coincides with the background on 
the disk, so $\mathbf{a}^{out}_{\text{\tiny R}}=\mathbf{a}_{\text{\tiny R}}=\cl$. To fix the background on the inner 
boundary, we extract the contribution due to the open string tachyon
\begin{equation}\label{pf-ann1s}
\int_0^1\frac{dq}{q}\,\bigl(-\ln q\bigr)^{\frac{d}{2}-1}
\ep^{-(\mathbf{a}_{\text{\tiny R}})^2-q(\mathbf{a}^{in}_{\text{\tiny R}})^2}\,
\ep^{-\frac{\pi^2}{\ln q}}
\end{equation}
and use the fact that $\mathbf{a}_{\text{\tiny R}}=\cl$. Thus we get for the background
\begin{equation}\label{inb}
\bigl(\mathbf{a}^{in}_{\text{\tiny R}}\bigr)^2=-\frac{1}{q}\mathbf{a}_{\text{\tiny R}}^2-
\frac{\pi^2}{q\ln q}\,\mathbf{a}_{\text{\tiny R}}^2
\quad.
\end{equation}
Note that in order to extract the contribution due to the open string tachyon one has to consider the $q\rightarrow 1$
limit that makes the right hand side of Eq.\eqref{inb} positive.

Finally, the partition function is given by 
\begin{equation}\label{pf-anns1}
\begin{split}
Z_1^{ann}(\mathbf{a}_{\text{\tiny R}})
=i\oh V_d \bigl(8\pi^3\ap)^{-\frac{d}{2}}\pi^4
\int_0^1
&\frac{dq}{q}\,\bigl(-\ln q\bigr)^{\frac{d}{2}-5}
\ep^{\frac{\pi^2}{\ln q}\mathbf{a}^2_{\text{\tiny R}}}\,
\prod_{n=1}^{\infty}\bigl(1-q^{2n}\bigr)^{-8}
\\
&\times\frac{1}{q}
\Bigl[\prod_{n=1}^{\infty}\bigl(1+q^{2n-1}\bigr)^8-
\prod_{n=1}^{\infty}\bigl(1-q^{2n-1}\bigr)^8\Bigr]
\quad.
\end{split}
\end{equation}

\subsubsection{Computations Using Cylinder}

Having computed the partition function on the annulus, let us see what happens on the cylinder. To this end, we 
start with the open NSR string on the strip and add the boundary interaction
\begin{equation}\label{bons}
S_b=\frac{1}{4}\int d\tau d\sigma \Bigl[T^2(X)\delta(\sigma)+T^2(X)\delta(\pi-\sigma)\Bigr]
\quad
\end{equation}
to the standard worldsheet action. As in the bosonic case, we consider a constant background field $T=a$. 

In computing the partition function on the strip, the only novelty compared to the bosonic computation 
is again that the renormalized coupling is now defined as $\ar=\sqrt{r}a$. This leads to a partition function 
$Z_0=\ep^{-\ar^2}$ that together with the definition \eqref{pkud} provides the same potential as in \eqref{ss} 
after the identification $\ar=\cl$. 

Let us now consider the partition function on a finite strip, the cylinder. Evaluating the path integral, we find
\begin{equation}\label{pfs1}
Z_1(\ar)=i\oh V_d\int_0^\infty\frac{dt}{t}\,\bigl(8\pi^2\ap t\bigr)^{-\frac{d}{2}}\ep^{-\pi t\ar^2}\,
\eta(it)^{-8}\Bigl[ f_{NS}^8(t)-f_R^8(t)\Bigr]
\quad,
\end{equation}
where $f_{NS}$ and $f_R$ come from the NS and R sectors, respectively. These have the product 
representations: 
\begin{equation*}
f_{NS}(t)=\ep^{\frac{\pi t}{24}}\prod_{n=1}^\infty\Bigl(1+\ep^{-\pi t(2n-1)}\Bigr)
\quad,\quad
f_{R}(t)=\sqrt{2}\,\ep^{-\frac{\pi t}{12}}\prod_{n=1}^\infty\Bigl(1+\ep^{-2\pi tn}\Bigr)
\quad.
\end{equation*}
As a consistency check, we again extract from $Z_1(\ar)$ the contribution due to the open string tachyon 
\begin{equation}\label{checks}
\int_0^\infty dt\,t^{-1-\frac{d}{2}}\,
\ep^{(1-\ar^2)\pi t}
\quad.
\end{equation}
A comparison with the field theory result of section 2 gives $\ar=\cl$. We have found 
the same relation between the renormalized value of the tachyon background and the classical field as at the tree level 
that is of course the desired result. Moreover, a simple algebra shows that the expressions for the partition function 
on the cylinder and annulus coincide. This is again the desired result. Note that we have not used the conformal 
map for mapping the annulus into the cylinder from the beginning. The reason for doing so is that we do not know 
the transformation law for the background fields.

As in the bosonic case, the partition function can be rewritten as a sum of the point-particle free energies 
(in the Coleman-Weinberg representation). It is natural to treat the leading term as the tachyon contribution and 
compare it with the field theory result. 

\subsubsection{Partition Function Analysis}

Having computed the partition function in terms of a loop of open string modes, one can formally analyze it by 
expanding the products in $e^{-\pi t}$. This yields
 \begin{equation}\label{os}
Z_1(\ar)=i\oh V_d\int_0^\infty\frac{dt}{t}\,\bigl(8\pi^2\ap t\bigr)^{-\frac{d}{2}}
\ep^{(1-\ar^2)\pi t}\,
\Bigl[1-8\,\ep^{-\pi t}+\dots\Bigr]
\quad.
\end{equation}
In the limit $t\rightarrow\infty$ a possible divergence can come only from the leading term if the absolute value 
of the background is less than its critical value $a_0=1$. As in the bosonic case, one can use the analytic 
continuation to fix the problem. To be more precise,
\begin{equation}\label{os1}
\int_0^\infty\,dt\,t^{-1-\frac{d}{2}}\ep^{-(\ar^2-1)\pi t}=
\ep^{i\nu\pi\frac{d}{2}}\,\pi^{\frac{d}{2}\,}\vert \ar^2-1\vert^{\frac{d}{2}}\,\Gamma(-\frac{d}{2})
\quad,
\end{equation}
where $\nu=0$ if $\vert\ar\vert >a_0$ and $\nu=\pm 1$ if $\vert\ar\vert <a_0$.

Now, we come to the question of whether the series over all open string massive modes is convergent or not. 
Unlike the bosonic case, here the sum is alternating, so there is a chance to get a finite result. We will 
not directly examine this problem but use our experience with the bosonic case where the divergency 
has been attributed to the closed string tachyon mode. Since there is no tachyon among the NS-NS states, we 
claim that the sum is convergent.

We can readily rewrite the products in \eqref{pfs1} in terms of $\ep^{-\frac{\pi}{t}}$. This yields 
another representation of the partition function. Expanding now in $\ep^{-\frac{\pi}{t}}$, we have
\begin{equation}\label{os3}
Z_1(\ar)=8i\,V_d\int_0^\infty\frac{dt}{t^{-3}}\,\bigl(8\pi^2\ap t\bigr)^{-\frac{d}{2}}
\ep^{-\ar^2\pi t}\,
\Bigl[1+16\,\ep^{-\frac{2\pi}{ t}}+\dots\Bigr]
\quad.
\end{equation}
As in section 3.1, the fact that the Feynman propagator of a scalar particle of mass $m$ in $d$ dimensions 
has the integral representation \eqref{fpr1} allows us to rewrite this expansion as
\begin{equation}\label{os4}
Z_1(\ar)=2i\pi\,V_d\,(4\pi^2\ap)^{4-d}
\Bigl[G_{10-d}\bigl(R;0\bigr)+16\,G_{10-d}\bigl(R;4/\ap\bigr)+\dots\Bigr]
\quad,
\end{equation}
where $R=\sqrt{2\pi^2\ap}\vert\ar\vert$. Note that in \eqref{os4} the closed string modes appear 
with the same masses as in the case $\ar=0$. 

Now let us check whether the sum over all massive closed string modes is convergent or not. To do so, we 
proceed in precisely the same way as in section 3.1. Using the asymptotics of the propagator \eqref{fpr3}, 
we get 
\begin{equation}\label{os5}
\sum_{n=1}^\infty c_{n}\,G_{10-d}\bigl(R;4n/\ap \bigr)
\sim
\sum_{n=1}^\infty c_n\,n^{\frac{7-d}{4}}\,
\ep^{-\sqrt{8\pi^2n}\,\vert\ar\vert}
\quad,
\end{equation}
where $c_n$ is defined as the coefficient of $q^n$ in $\prod_{n=1}^\infty\Bigl(\frac{1+q^n}{1-q^n}\Bigr)^8$. 
A similar technique as it is used to estimate $d_n$ for large $n$  can be used to get the asymptotics for 
$c_n$. The result is simply \cite{gsw}
\begin{equation}\label{c}
c_n\sim n^{-\frac{11}{4}}\,\ep^{\sqrt{8\pi^2n}}
\quad.
\end{equation}
Comparing the exponent to that of \eqref{os5}, we see that the sum over massive modes is convergent if 
$\vert\ar\vert>1$. In other words, the absolute value of the background field must be larger than its critical 
value $a_0$ we have found by studying the contribution to the partition function of the open string tachyon.

It should be also noted that a simple check shows that Eq.\eqref{os5} leads to the same scaling behavior 
near the critical point as Eq.\eqref{os1} does. 

It is interesting to consider the large $\ar$ limit again. Since there is no closed string tachyon, the asymptotic of 
the partition function is due to the massless NS-NS modes. Explicitly, 
\begin{equation}\label{os6}
Z_1(\ar)=i\frac{2^{3-d}}{\pi^4}\,V_d\,(2\pi\ap)^{-d/2}\Gamma(4-d/2)\,
\vert\ar\vert^{d-8}
\quad.
\end{equation}
Note that the right hand side of \eqref{os6} is singular in the two cases $d=8$ and $d=10$. We postpone discussion of 
this complication until section 4.

It is of some interest to determine the real part of $Z_1$ which appears if the absolute 
value of the background field is less than the critical one. For doing so, we use  the representation of the partition 
function as a loop of open string modes. From \eqref{os1} we get
\begin{equation}\label{os7}
\Re Z_1(\ar)=\pm\frac{\pi}{2}V_d\,(8\pi\ap)^{-d/2}
\frac{\bigl(1-\ar^2\bigr)^{\frac{d}{2}}}{\Gamma(1+d/2)}H(1-\ar^2)
\quad.
\end{equation}
%_____________________________________________________________________________

To summarize, we have computed the partition functions by evaluating the corresponding path integrals. It 
turns out that the results admit the representation as a sum of the point-particle free energies (in the 
Coleman-Weinberg representation) over the open string spectrum but with the shifted masses of the particles. This 
allows us to compare the tachyon contribution with the corresponding field theory result \footnote{In the context of 
brane-antibrane systems, this fact has also been noted in \cite{al}. }. Moreover, the partition functions are easily 
rewritten via closed string modes with the standard masses. This reveals the transverse directions to the brane 
and looks like closed string modes appearing from one brane, propagating a distance $R$, and then disappearing 
on the other brane. In other words, the description in terms of closed strings assumes that the original brane splits 
into two parallel branes separated by a distance depending on the vev of the open string tachyon. Note that 
the expressions for the partition functions are similar to those for the partition functions of open strings stretched 
between two parallel branes (see, e.g., \cite{dj}). However, our situation is absolutely different as we are interested 
in the partition functions of open string modes on a single brane but in the presence of a constant tachyon background. 
Thus, it is a priori unclear why in this case closed strings ``see'' the background in such a way.
%______________________________________________________________________________
\section{Attempt of Synthesis}
\renewcommand{\theequation}{4.\arabic{equation}}
\setcounter{equation}{0}
By now we have considered the one-loop corrections to the tree level potentials as they follow from some 
field theory models and the partition functions of open strings in the presence of a constant tachyon 
background. The remaining question is how to properly combine them in order to find the one-loop 
corrections to the tachyon potentials.

%____________________________________________________________________________
\subsection{Effective Potential at One Loop}
As we have already mentioned, in the case of a constant tachyon background the partition functions can be rewritten as 
the sums of the point-particle contributions over the open string spectrum. It follows from this fact that we already have 
the desired correspondence with the field theory results on the level of the partition functions. Thus it seems natural, modulo 
terms due to  a normalization and renormalization conditions (see \eqref{p1}), to identify the one-loop 
correction to the effective tachyon potential with the open string partition function in both cases, the bosonic and 
supersymmetric ones. Explicitly, 
\begin{equation}\label{s1}
V_{1-loop}=\frac{i}{V_d} Z_1
\quad.
\end{equation}
This also means the identification of the renormalized background field $\ar$ in $Z_1$ with the 
classical field $\cl$.  

A natural question to pose at this point is whether there are other indications in favor of this identification. This time 
we have one which is as follows. An interesting point of view on the formulae
\eqref{wd} and \eqref{pkud} proposed in \cite{kud1} is to relate them with a possible way of subtraction of the 
M\" obius infinities from the disk partition functions. Indeed, it is well known that the M\" obius group volume 
is linearly divergent in the bosonic case while it is finite in the supersymmetric case \cite{mob}. It is easy to see 
that the definition \eqref{wd} leads exactly to the cancellation of the divergency. As there is no need to deal with 
the divergency in the supersymmetric case, it now seems natural to identify the partition function with the 
effective action as in \eqref{pkud}. Having the scenario for the disk, it is a simple matter to generalize it for 
all higher genera. Since in this case there is no M\" obius invariance \footnote{We do not assume the use of 
extended parameterizations for moduli of higher genus surfaces as, for example, the Schottky parameterization, 
where the SL(2) invariance is present for any genus.}, there is no need to deal with the M\" obius infinities and, as a 
consequence, one can simply identify the corresponding correction to the action with the partition function, namely
\begin{equation}\label{s2}
S_n=Z_n
\quad,\quad
\text{for}
\quad
n\leq 1
\quad.
\end{equation}
From this point of view, Eq.\eqref{s1} is a special case of this relation. Note that for $n=1$ this relation 
has already been suggested in the literature as a possible ansatz for finding the one-loop effective actions 
via the corresponding open string partition functions \cite{1b,1s}. Moreover, in the context of the effective 
action for massless modes it was proposed in \cite{aat}. 

Having combined the worldsheet constructions with the field theory results, we have the expressions for the 
one-loop corrections to the effective potentials. So, it is time to consider them in more detail. To proceed, we 
will specialize first to the case of the NSR string as it turns out to be simpler than the case of the bosonic string.

\subsubsection{The Supersymmetric Case}
As we have seen, the partition function has a real part if the absolute value of the background field is less than 
its critical value. As a result, this will cause the one-loop correction to the potential to be complex for this range. 
This difficulty is due to the open string tachyon, or equivalently closed massive modes. It is advantageous to analyze it 
in terms of the open string tachyon as it allows us to use the well developed machinery for analyzing perturbative 
instabilities in field theories with nonconvex potentials (see, e.g., \cite{ww}). Thus, a physically meaningful quantity 
to compute in this range is 
the decay rate per unit volume of an initial unstable state. Formally, in terms of the imaginary part of 
$V_{1-loop}$ it is given by 
\begin{equation}\label{dr}
\Gamma=2\,\Im V_{1-loop}
\quad.
\end{equation}
Since this quantity must be positive, this requirement fixes the sign in our formula \eqref{os1} for the 
analytic continuation. Then from \eqref{os7} together with \eqref{s1} it follows that 
\begin{equation}\label{dr1}
\Gamma=\pi(8\pi\ap)^{-d/2}
\frac{\bigl(1-\cl^2\bigr)^{\frac{d}{2}}}{\Gamma(1+d/2)}H(1-\cl^2)
\quad.
\end{equation}

On the other hand, for $\vert\cl\vert >1$ there is no problem with the imaginary part. This allows us to write 
down the one-loop correction to the tachyon potential
 \begin{equation}\label{sp}
V_{1-loop}(\cl)=-\oh \int_0^\infty\frac{dt}{t}\,\bigl(8\pi^2\ap t\bigr)^{-\frac{d}{2}}\ep^{-\pi t\cl^2}\,
\eta(it)^{-8}\Bigl[ f_{NS}^8(t)-f_R^8(t)\Bigr]
\quad.
\end{equation}
At this point, let us make a remark. From our discussion in section 3 it follows that $V_{1-loop}$ is finite, so 
$V_{1-loop}(\infty)=0$. Thus, the use of $\text{m}^2_0=\infty$ in Eq.\eqref{p1} or $\cl=\infty$ as a normalization 
of $V_{1-loop}$ does not lead to a modification of our results. To be precise, this is the case for $d<8$. We will 
return to this point shortly.

It is of some interest for what follows to look at the one-loop correction in the limit of large $\cl$. Using the results
for the asymptotical behavior of the corresponding partition functions, we immediately get from relation \eqref{s1}
\begin{equation}\label{spl}
V_{1-loop}(\cl)=-\frac{2^{3-d}}{\pi^4}\,(2\pi\ap)^{-d/2}\Gamma(4-d/2)\,
\vert\cl\vert^{d-8}
\quad.
\end{equation}

A special consideration is needed for $d\geq 8$ where the argument in the $\Gamma$-function becomes 
nonpositive. It might lead to divergencies as the $\Gamma$-function has poles. In fact, there is no problem 
for $d=9$, as in this case we can use the familiar analytic continuation for the $\Gamma$-function. As a result, 
we end up with
\begin{equation}\label{sp2}
\frac{\pi}{\sqrt{2}}\bigl(4\pi^2\ap\bigr)^{-9/2}\,\vert\cl\vert
\quad.
\end{equation}
However, this procedure does not work for $d=8, 10$. The problem is that the propagators of massless 
particles are ill-defined  in the corresponding codimensions ($2$ and $0$). First, let us consider the case $d=8$. 
As is known, one way to define the propagator of a massless scalar particle in two dimensions is to introduce a 
long-distance cutoff (scale). Then the propagator takes the form $\ln R/L$. Note that a constant part can be fixed 
by a proper normalization at $R=L$. Keeping this in mind, we proceed in the following way. We use dimensional 
regularization for expression \eqref{spl} 
\begin{equation}\label{d=8}
\bigl(4\pi^2\ap\bigr)^{-4}\Bigl[\frac{1}{\epsilon}+\ln\vert\cl\vert+const\Bigr]
\quad.
\end{equation}
However, to emphasize the physical role of the cutoff, we exchange $1/\epsilon$ for $\ln T_n$ and fix a constant part 
by a normalization at $\cl=T_n$. Then this expression takes the form
\begin{equation}\label{sp3}
\oh\bigl(4\pi^2\ap\bigr)^{-4}\ln\Bigl(\frac{\cl}{T_n}\Bigr)^2
\quad.
\end{equation}

Although for $d=10$ there is no direct analogy with field theory (this is the case of a spacefilling brane, so 
the corresponding codimension is zero), the preceding computation formally generalizes without difficulty to this case. 
The result is given by 
\begin{equation}\label{sp4}
-\frac{\pi}{4}\bigl(4\pi^2\ap\bigr)^{-5}\,\cl^2\,
\ln\Bigl(\frac{\cl}{T_n}\Bigr)^2
\quad.
\end{equation}

At this point, there is a subtle issue to discuss. We can evaluate the propagators by cutting off the upper limit of 
integration in expression \eqref{fpr1} for $m^2=0$. In particular, a simple algebra shows that in the case of a 
spacefilling brane the corresponding integral is $\int^L dl\,\ep^{-\frac{R^2}{4l}}\sim L-\frac{1}{4}R^2\ln L$. 
Here we restrict ourselves to the leading asymptotics in $L$. In fact, the power like divergency is well-known in string 
theory without tachyon background ($R=0$). It is removable by the Fischler-Susskind mechanism \cite{fs} which 
assumes that $L$ is identified with the worldsheet cutoff $\ln\mu$. This just provides a correction to the worldsheet beta 
functions. For $R>0$, the situation is quite different. As we have seen, it is possible to avoid power-like divergencies 
by using dimensional regularization. However, the novelty now is the appearance of logarithms. Note that a naive 
attempt to apply the Fischler-Susskind mechanism to the logarithms runs into trouble because it would lead to 
worldsheet divergencies like $\ln\ln\mu$ \footnote{Another problem with the Fischler-Susskind mechanism in the 
case of a nonzero tachyon background was noticed by Craps, Kraus, and Larsen \cite{1b}. They found a singularity that 
can not be interpreted as being due to closed string modes exchange. We have no such problem as in our case the 
factorization in terms of closed string modes is manifest (see \eqref{inf3} and \eqref{os4}).}. To interpret these 
logarithms, let us look at the representation of the partition function via open string states. It looks like a sum of the 
point-particle free energies written in the Coleman-Weinberg form (see, e.g., \eqref{particle}). For large values of 
$\cl$, the integrals are generically divergent at the lower limit that corresponds to short-distance divergencies. 
The use of dimensional regularization (see, e.g., \eqref{int}) shows a pole $1/\epsilon$ for even $d$ which can be 
exchanged for a logarithm. Note that this is the ultraviolet divergency of the theory on a brane. To get the representation 
via closed string states, we changed variables as $l\rightarrow 1/l$. It is clear that it results in the logarithmic 
divergencies at the upper limit of integration. But this is exactly what we found in the closed string channel. Moreover, 
we found that closed string states propagate in the transverse directions to the brane i.e., in the bulk. Thus  the moral 
of this story in fact is known: the UV/IR correspondence. Note that the real situation is more subtle than what 
we sketched above. Indeed, the logarithms appear only for codimensions $0$ and $2$ while the Coleman-Weinberg 
formula (see \eqref{int}) assumes their appearance for any even dimension. An answer to this is the cancellation 
between contributions of different open string states. It is clear that further work is needed to make this more rigorous. 

As we have already mentioned, the relation \eqref{s1} is defined modulo a normalization and renormalization conditions 
(see \eqref{p1}). From our analysis of the partition functions it follows that the refinement is needed 
only for $d\geq 8$. We propose to normalize the corrections for these cases at some large $\cl=T_n$. As a result, the 
one-loop corrections at large $\cl$ are given by 
\begin{equation}\label{spec}
V_{1-loop}(\cl)=
\begin{cases}
\oh\bigl(4\pi^2\ap\bigr)^{-4}\ln\Bigl(\frac{\cl}{T_n}\Bigr)^2
&\quad\text{if}\quad d=8\,\,,\\
\frac{\pi}{\sqrt{2}}\bigl(4\pi^2\ap\bigr)^{-\frac{9}{2}}\,\bigl(\vert\cl\vert-T_n\bigr)
&\quad\text{if}\quad d=9\,\,,\\
-\frac{\pi}{4}\bigl(4\pi^2\ap\bigr)^{-5}\,\cl^2\,
\ln\Bigl(\frac{\cl}{T_n}\Bigr)^2
&\quad\text{if}\quad d=10\,\,.
\end{cases}
\end{equation}

The main conclusion we can draw from the large $\cl$ behavior of the one-loop corrections is that except for a special 
case at $d=10$ they are not falling at infinity. This will have a strong impact on our discussion of tachyon condensation.

Now we have all at our disposal to see how the one-loop approximation to the tachyon potential looks like. A typical 
picture is that of figure 1 \footnote{For the sake of simplicity, we consider the case of one new local minimum 
$T_\ast$. We also assume that the vev of the dilaton is large enough to discard the one-loop corrections near 
$\vert\cl\vert=1$.}. There are two different scenarios that crucially dependent on the string coupling constant 
near the perturbative vacuum ($\cl=0$):

\noindent I. The strong coupling regime ($\lambda_e\sim1$) appears before $\cl$ becomes large. This means 
that we gain nothing essentially new by the one-loop corrections as the system rolls down into the strong coupling 
regime as it also does in the tree level approximation.

\noindent II. The strong coupling regime appears at a very large value of $\cl$, where our approximation 
for the propagator \eqref{fpr3} is valid \footnote{Clearly, this requires a larger vev of the dilaton.}. In this case, a 
new minimum $T_\ast$ appears at a finite value of the tachyon field \footnote{Note that $d=10$  is a special 
case. We will return to this point in section 5.}. So the one-loop corrections become relevant and, as a 
result, the system does not roll down into the strong coupling regime as the tachyon condenses. This means 
that the phenomenon of tachyon condensation can be described in this case by weakly coupled theory.
%________________________  Fig - 1  __________________________________
%
\vspace{.3cm}
\begin{figure}[ht]%[htbp]
\begin{center}
 \includegraphics{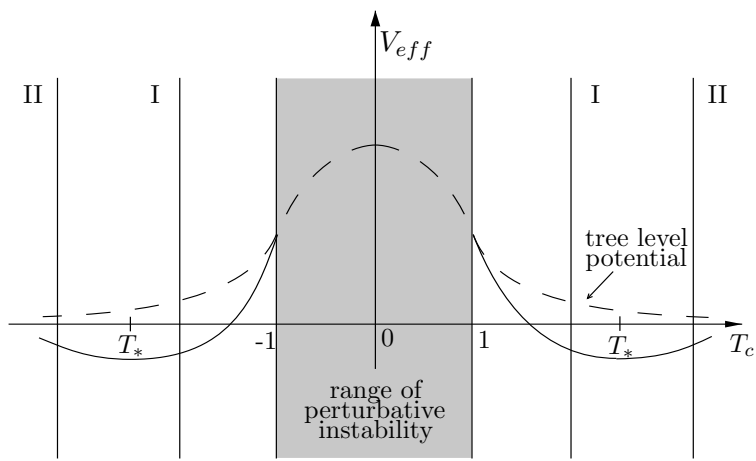}
 \caption{One-loop corrected tachyon potential in NSR string.}
    %%\label{fig:graph}
 \end{center}
\end{figure}
%_____________________________________________________________________
\subsubsection{The Bosonic Case}
The situation in the bosonic case is much more involved. Indeed, as we have seen in section 3.1, there are 
two contributions to $Z_1$ (see \eqref{rez1} and \eqref{rez2}) that become relevant in studying the imaginary 
part of the one-loop potential. Following our experience with the supersymmetric case, we can easily elaborate 
the contribution due to the open string tachyon. Thus, we interpret it again as the perturbative field theoretical 
instability and its contribution to the decay rate per unit volume of an initial 
unstable state is simply
\begin{equation}\label{drop}
\Gamma^{(op.tach.)}=\pi(8\pi\ap)^{-d/2}
\frac{\bigl(2-\cl\bigr)^{\frac{d}{2}}}{\Gamma(1+d/2)}H(2-\cl)
\end{equation}
that is in agreement with the results that are already avaliable in the literature for the case of $\cl=0$ (see \cite{1b}). 
However, the other contribution to the imaginary part of the one-loop correction 
makes the situation more involved. Remind that it can not be interpreted in 
terms of the open string tachyon, as it comes from open string massive modes or, equivalently, the closed string 
tachyon. In practice, the last turns out to be more useful. Formally, we have 
\begin{equation}\label{clos}
V_{1-loop}^{(cl.tach.)}=
-(4\pi^2\ap)^{12-d}\,\frac{\pi}{2^{11}}\,\Im\,G_{26-d}\bigl(\sqrt{2\pi^2\ap\cl};-4/\ap\bigr)
\quad
\end{equation}
that as it follows from \eqref{imr} in the limit of large $\cl$ reduces to
\begin{equation}\label{clos1}
V_{1-loop}^{(cl.tach.)}\sim\mp\cl^{\frac{d-25}{4}}\sin\Bigl[\pi\bigl(\sqrt{8\cl}+\frac{3-d}{4}\bigr)\Bigr]
\quad.
\end{equation}
A naive attempt to make use of the above formula for analyzing a possible instability at $\cl=\infty$ fails. Indeed, as 
in the supersymmetric case, we can consider two scenarios. The point is that for both of them $\cl=\infty$ 
belongs to the strong coupling regime, so this is outside the reach of our approach.

Anyway, we have found that in the problem of tachyon condensation, there is the instability due to the closed string 
tachyon. The physical meaning of this effect should be further clarified.

Note also that there was a speculation with weak evidence for it in the literature \cite{kud} that interprets the 
instability of the vacuum $\cl=\infty$ as the closed string tachyon instability which leads to the true vacuum 
at $\cl=-\infty$. We will return to this issue in section 5.
%____________________________________________________________________________
\subsection{Dynamics of Tachyon Condensation}
Given what was said in the previous sections, the interpretation of the dynamics of tachyon condensation in 
terms of closed strings is clear. Indeed, initially a D-brane is at the open string perturbative vacuum $\cl=0$, i.e. 
at a local maximum of the potential. Then as the tachyon is rolling down, the brane is getting split into two 
branes with a distance in between depending on $\cl$, as shown on figure 2.
%________________________  Fig - 2 __________________________________
%
\vspace{.3cm}
\begin{figure}[ht]%[htbp]
\begin{center}
 \includegraphics{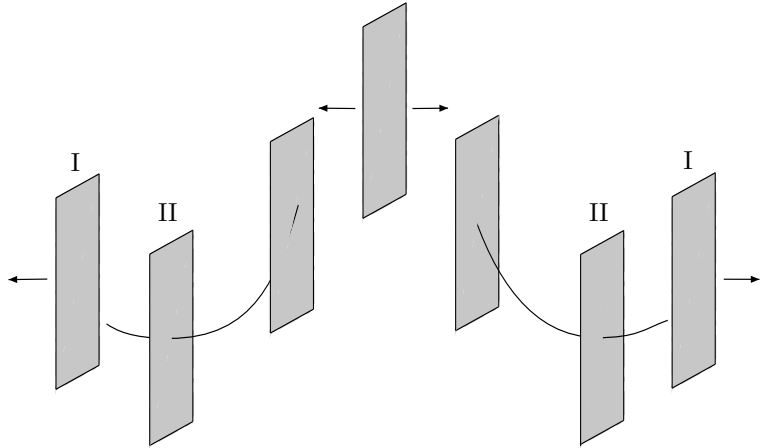}
 \caption{Dynamics of tachyon condensation. A part of the profile of the potential for the 
supersymmetric case is also shown.}
 %%\label{fig:graph}
 \end{center}
\end{figure}
%_____________________________________________________________________

The further behavior depends on details: in the bosonic case, there is the instability due to the closed string 
tachyon and the development of this instability is not clear yet; in the supersymmetric case, the further behavior 
crucially depends on the chosen scenario. For scenario I, the branes enter the strong coupling regime where 
our tools become inappropriate to explore the details of the process of condensation. For scenario II, the branes 
are getting stabilized at some distance as the potential develops the new minimum at $T_\ast$.  

It is of some interest to notice that there is an analogy with field theory models that undergo spontaneous symmetry 
breaking \cite{linde} that helps to realize the brane dynamics of figure 2 at the beginning of the decay. Indeed, 
given an effective potential, for example $V_{eff}\sim\bigl(T^2-T_\ast^2\bigr)^2$, one can study the dynamics 
of spontaneous symmetry breaking by setting an initial narrow distribution of the scalar field at the 
maximum of the potential ($T=0$). Then, by quantum fluctuations it spreads out and shows two maxima. 
To reveal such a picture in the problem of interest, let us recall two key facts we have already seen. The first is 
that in terms of closed strings $\cl$ transforms into a transverse direction to a brane. It is the reason why we 
have drawn a part of the profile of the potential in figure 2. The second is that a brane itself can be considered 
as a lump solution of the higher dimensional worldvolume theory that allows one to say that this solution plays a role
of the initial distribution in field theory. Finally, let us note that such a description of the decay shows that these 
lumps interact by the closed string exchanges. 

%_____________________________________________________________________
\section{Concluding Comments}
\renewcommand{\theequation}{5.\arabic{equation}}
\setcounter{equation}{0}
What we have learned is  the one-loop approximation to the tachyon potentials. We did so by the worldsheet 
methods. The key fact to stress at this point is that the tachyon backgrounds we have considered are not due to 
conformal primary fields. This might seem like a problem as it would lead to an ambiguity. To fix this problem, we 
have used the correspondence with field theory. In the bosonic case, our analysis has also revealed the instability of 
D-branes due to the closed string tachyon. Moreover, it allowed us to see the dynamics of tachyon condensation 
in terms of closed strings that is similar to the dynamics of spontaneous symmetry breaking in field theory.

There are many issues that deserve to be further clarified. Let us mention some of them that are seemed the 
most important to us.

The first is the physical meaning of the instability of bosonic D-branes due to the closed string tachyon. A proposal 
of \cite{kud} that relates it to the tunneling through the potential barrier from $\cl=\infty$ to $\cl=-\infty$ seems 
inappropriate in our case. Indeed, the tunneling results in the nonperturbative instability that behaves 
like $\exp\bigl(-1/\lambda_e\bigr)$. Clearly, we do not have such a behavior as we have found the closed string 
tachyon instability which is perturbative. Rather, this instability can be attributed to a peculiarity of euclidean 
field theory with the tachyonic particle.

The second issue is how to incorporate D-instantons. It is obvious that the field theory analysis of section 2 
does not help in doing so because $d=0$ for this case. In fact, the problem is deeper: Even at the tree level 
the worldsheet methods, e.g. BIOSFT, have turned out to be useful to 
check some of the Sen conjectures about tachyon condensation. But they have stumbled on D-instantons as 
it is unclear how to proceed further namely, how to get the closed string perturbative vacuum without 
any brane excitations. It seems that at this point the worldsheet methods should be revised as the appearance of
D-instantons usually requires a modified construction of string perturbation theory (see, e.g., \cite{djD}). 
Note also that a formal substitution $d=0$ into formulae of section 4, e.g. Eq.\eqref{s1}, is not legitimate as our 
analysis is based on the use of the euclidean metric for the transverse directions to branes rather than the 
Minkowski metric. This is another subtlety we face with D-instantons.

As we have seen, the definition of the one-loop correction $V_{1-loop}$ to the tachyon potential through the open 
string partition function $Z_1$ (see \eqref{s1}) results in a potential which is complex for some range of $\cl$. It is 
known that in this case the physically meaningful quantity to compute is the decay rate per unit volume which is 
defined in terms of the imaginary part of $V_{1-loop}$. It was suggested in \cite{ww} that in field theory models 
the real part of the potential can, nevertheless, be interpreted as the energy density. This assumes the use of an analytic 
continuation of $V_{1-loop}$ to dangerous regions. It is clear that an attempt to do so in the case of the bosonic string 
fails as there is no region where $V_{1-loop}$ is real. In the supersymmetric case the situation is better since 
the one-loop correction is real for $\vert\cl\vert \geq 1$. So, it might be possible to define the energy density 
for $\vert\cl\vert <1$ as well. As an example, let us briefly illustrate this for odd $d$. The results of sections 3 and 4 
yield
\begin{equation}\label{ancon}
V_{1-loop}=-\oh\bigl(8\pi\ap\bigr)^{-\frac{d}{2}}\Gamma(-\frac{d}{2})
\sum_{n=0}^\infty b_n\bigl(n+\cl^2-1\bigr)^{\frac{d}{2}}
\quad,
\end{equation}
where $b_n$ is defined as the coefficient of $q^n$ in 
$\prod_{n=1}^\infty \Bigl(\frac{1-q^{2n-1}}{1-q^{2n}}\bigr)^8$. As we have noticed in section 3, the sum 
is convergent, so the only problem is that it becomes complex valued 
for $\vert\cl\vert<1$ due to the contribution from the open string tachyon. We use the analytic 
continuation (see \eqref{os1}) to obtain the one-loop correction for this region
\begin{equation}\label{ancon1}
V_{1-loop}=-\oh\bigl(8\pi\ap\bigr)^{-\frac{d}{2}}\Gamma(-\frac{d}{2})
\sum_{n=1}^\infty b_n\bigl(n+\cl^2-1\bigr)^{\frac{d}{2}}
\quad.
\end{equation}
It is interesting to note that our result simply assumes dropping the contribution of the tachyon. Moreover, the tree level 
vacuum $\cl=0$ remains the extremum at the one-loop level too \footnote{By differentiating \eqref{ancon1}, one 
makes the sum more convergent than it was before.}. The latter is also clear by applying a general argument:
Since $V_{1-loop}$ is an even function of $\cl$, its derivative is odd. The case of even $d$ is more delicate as it 
requires the treatment of the divergencies. However, if the symmetry $\cl\rightarrow -\cl$ is  preserved then the above 
argument also holds. 

The case of a spacefilling brane is quite intricate. Indeed, in this case our interpretation of a brane decay as 
the splitting of the original brane into two parallel branes becomes problematic. Moreover, the logarithmic 
divergency obviously deserves the better understanding. 

Finally, the analogy of field theory models with spontaneous symmetry breaking deserves to be studied further. 
Let us note at this point that the idea to use a Higgs like mechanism within the process of tachyon condensation 
has been already proposed in the literature \cite{gera1}. However, our discussion of it is quite different.

%__________________       Acknowledgments   ________________________
\vspace{.35cm} {\bf Acknowledgments}

\vspace{.25cm} 
We would like to thank J. Polchinski, A. Sen, S. Shenker, W. Taylor for useful discussions and also 
H. Dorn and  A.A. Tseytlin for comments and reading the manuscript. Part of this work was done while O.A. was 
participating in M-theory program at the Institute of Theoretical Physics at UCSB supported by the NSF under 
grant No. PHY99-07949. This work is also supported in part by DFG under grant No. DO 447/3-1 and the 
European Commission RTN Programme HPRN-CT-2000-00131. T.O. would like to thank the 
Graduiertenkolleg ``The Standard Model of Particle Physics - Structure, Precision Tests and Extensions'',
maintained by the DFG.

%_________________________  Appendix - A ______________________________

\vspace{.25cm} {\bf Appendix}
\renewcommand{\theequation}{A.\arabic{equation}}
\setcounter{equation}{0}

\vspace{.25cm} 
In this appendix we will briefly discuss the propagator of the tachyon of mass $im$ in $d$-dimensional 
euclidean space based on the integral representation \eqref{fpr}. For what follows, it is convenient to take such a 
frame with $x'=0$ and $px=p_1x^1$.

To gain intuition on what is going on, let us begin with $d=1$. In this case, the integral suffers from divergencies 
at $p_1=\pm m$. A possible way to treat them is to slightly deform the integration 
contour in their vicinity. Obviously, there are four possibilities in doing so:
%________________________  Fig - 3 __________________________________
%
\vspace{.3cm}
\begin{figure}[ht]%[htbp]
\begin{center}
 \includegraphics{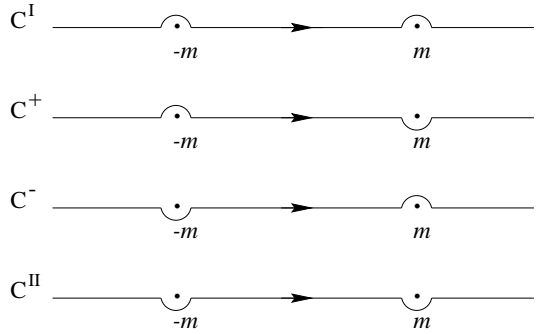}
 \caption{Contours used in the definition of the propagator.}
    %%\label{fig:graph}
 \end{center}
\end{figure}
%_____________________________________________________________________

After closing the contours at infinity \footnote{We assume that $x^1>0$.}, the integrals are given by the 
residue of the integrands. So, we get 
\begin{equation}\label{d1}
G^{\text I}_1(x;-m^2)=0
\quad,\quad
G^{\pm}_1(x;-m^2)=\pm\frac{i}{2m}\ep^{\pm imx}
\quad,\quad
G^{\text{II}}_1(x;-m^2)=-\frac{1}{m}\sin mx
\quad.
\end{equation}
Let us also note at this point that there is no ambiguity for a real mass $m$ because in this case the poles lie on the 
imaginary axis. Moreover, the closing of the contour at infinity picks up a contribution only from one pole.

To see ambiguity for generic $d$, we note that in our special frame the propagator admits the following 
representation
\begin{equation}\label{d2}
G_d(x;-m^2)=\frac{1}{(2\pi)^{d-1}}
\biggl[\int _{p^2_\perp>m^2}d^{d-1}p\,\,G_1(x; p^2_\perp-m^2)+
\int _{p^2_\perp<m^2}d^{d-1}p\,\,G_1(x; -m^2+p^2_\perp)
\biggr]
\quad,
\end{equation}
where $p^2_\perp=p_2^2+\dots +p_d^2$. Together with \eqref{d1}, this implies that the propagator is ambiguous. 

In fact, only $\text{C}^\pm$ are relevant for our discussion of the bosonic partition function. Indeed, the closed string 
tachyon contribution to the partition function \eqref{inf3} can be obtained modulo a numerical coefficient from the 
contribution of the lightest massive states by $m\rightarrow\pm im$. This is equivalent to the use of 
$\text{C}^\pm$ for the propagator because only in this case one pole contributes. In particular, 
the imaginary part of the propagator behaves at large $R$ as 
\begin{equation}\label{imr}
\Im ^{\pm}\,G_d(R;-m^2)=\pm\frac{1}{2\vert m\vert}\biggl(\frac{2\pi R}{\vert m\vert}\biggr)^{\frac{1-d}{2}}
\sin\Bigl[\vert m\vert R+\pi\,\frac{3-d}{4}\Bigr]
\end{equation}
as it follows from Eq.\eqref{fpr3} after the continuation to an imaginary mass. 

%__________________                      R E F S                    ______________________

%__________________________________________________________

\end{document}